\documentclass[aps,pra,reprint,amsmath,amssymb,showpacs]{revtex4-1}
\usepackage{bm,graphicx}
\newcommand*\ii{\mathrm{i}}
\newcommand*\ee{\mathrm{e}}

\begin{document}

\title{State Flip at Exceptional Points in Atomic Spectra}
\author{Henri Menke}
\email{Henri.Menke@itp1.uni-stuttgart.de}
\author{Marcel Klett}
\email{Marcel.Klett@itp1.uni-stuttgart.de}
\author{Holger Cartarius}
\author{J\"org Main}
\author{G\"unter Wunner}

\affiliation{
  Institut f\"ur Theoretische Physik 1,
  Universit\"at Stuttgart,
  70550 Stuttgart, Germany}


\begin{abstract}
  We study the behavior of the non-adiabatic population transfer between
  resonances at an exceptional point in the spectrum of the hydrogen atom.
  It is known that, when the exceptional point is
  encircled, the system always ends up in the same state, independent
  of the initial occupation within the two-dimensional subspace spanned by
  the states coalescing at the exceptional point. We verify this behavior for a realistic quantum system, viz.\
  the hydrogen atom in crossed electric and magnetic fields. It is also shown
  that the non-adiabatic hypothesis can be violated when resonances in the
  vicinity are taken into account. In addition, we study the non-adiabatic
  population transfer in the case of a third-order exceptional point, in which
  three resonances are involved.
\end{abstract}

\maketitle

\section{Introduction}

In many cases a very effective way of investigating open quantum systems with
reasonable effort, in particular to avoid expensive time-dependent
calculations, is made possible by non-Hermitian Hamiltonians
\cite{Moiseyev2011a}. A typical example are resonances, i.e.\ decaying unbound
states. With appropriate methods as the complex scaling approach
\cite{Reinhardt1982,Moiseyev1998,Moiseyev2011a,Delande1991,Ho1983} resonances
can be uncovered in a time-independent calculation as complex eigenvalues of
the stationary Schr\"odinger equation. It is well known that resonances show
characteristic effects not observable in Hermitian quantum systems. This is
in particular true close to exceptional points \cite{Heiss2012a,Hei99,kato,%
Moiseyev2011a}, isolated points in a physical parameter space, at
which two or even more eigenstates coalesce.

The appearance of exceptional points has theoretically been shown in unstable
lasers \cite{Berry2003}, optical wave guides \cite{Klaiman08a} and resonators
\cite{Wiersig2008,Wiersig2014a}. In quantum systems their existence has been
proved in atomic \cite{Magunov1999a,Magunov2001a,Latinne1995,cartarius_prl}
or molecular \cite{Lefebvre2009} spectra, in the scattering of particles at
potential barriers \cite{Hernandez2006}, in atom waves
\cite{Rapedius2010,Cartarius08a,Gutoehrlein13a,Abt2015a}, and in non-Hermitian
Bose-Hubbard models \cite{Graefe08a}. Their relation to Fano resonances
has been pointed out \cite{Magunov2003a,Heiss2014a,Schwarz2015a}. The experimental
verification of their physical nature was achieved in microwave cavities
\cite{Dembowski2001,Dietz2011,Bittner2014a}, electronic circuits \cite{Stehmann2004a},
metamaterials \cite{Lawrence2014a}, a photonic crystal slab
\cite{Zhen2015a}, and exciton-polariton resonances \cite{Gao2015a}.

A simple example is the two-dimensional matrix
\begin{equation}
  \bm M(\lambda) =
  \begin{pmatrix}
    1 & \lambda \\
    \lambda & -1 \\
  \end{pmatrix}
\end{equation}
with a one-dimensional complex parameter $\lambda$. The eigenvalues are given
by $\epsilon_1=\sqrt{1+\lambda^2}$ and $\epsilon_2=-\sqrt{1+\lambda^2}$. It is
obvious that these eigenvalues share a common degeneracy for
$\lambda=\pm\ii$ and the same holds for the eigenvectors, i.e.\ $\lambda=
\lambda_c=\pm\ii$ is an exceptional point. The example demonstrates one of
the most striking features of an exceptional point, which can be seen from
a power series expansion of the eigenvalues for a circle $\lambda(\varphi) 
= \ii + \varrho \ee^{\ii \varphi}$ with a small radius $\varrho$ around an
exceptional point,
\begin{equation}
    \epsilon_1 = \sqrt{2 \varrho} \ee^{\ii (\pi/4 + \varphi/2)} \; , \quad
    \epsilon_2 = \sqrt{2 \varrho} \ee^{\ii (5\pi/4 + \varphi/2)} \; .
\end{equation}
Evidently the eigenvalues interchange their position in energy space when the
exceptional point is encircled in a closed loop, i.e.\ $\varphi = 0\dots
2\pi$. Only after two circles the eigenvalues return to their original
positions, but the eigenvectors pick up a geometric phase, which is expressed
by a sign change, e.g.\ $[\psi_1,\psi_2] \xrightarrow{\smash{\text{circle}}}
[\psi_2,-\psi_1]$.

This is an example of a second-order exceptional point (EP2),
but also higher-order exceptional points are possible 
\cite{Heiss2012a}. The simplest extension is a
third-order exceptional point (EP3), at which three resonances coalesce,
i.e.\ have identical eigenvalues and eigenvectors \cite{Heiss2008a,%
  cartarius_pra,Cartarius08a,Demange2012a,Gutoehrlein13a,Eleuch2015a}.

\citet{uzdin} as well as \citet{Berry2011} have shown that the
adiabatic exchange of the states mentioned above will not be
observable for the true temporal evolution of an occupied resonance
state. Only one of the states behaves according to the adiabatic
expectation. The reason is that non-adiabatic effects can never be
neglected in the decay dynamics of resonances and the adiabatic
connections \cite{Arnold1995a} are no longer fulfilled. When an
exceptional point is encircled the occupation always ends up in the
same state, independent of the initial occupation within the
two-dimensional subspace of the resonances forming the exceptional
point
\cite{uzdin,Berry2011,Gilary2012,Gilary2013a,Greaefe2013a,Viennot2014a}. Recently
a careful analysis of the dynamics revealed that it is strongly
non-intuitive \cite{Milburn2015a}. Note that the adiabatic expectation
with the exchange of two resonances can always be extracted from
evaluations of their complex eigenvalues if the physical parameters
are changed in small steps on a closed loop in the parameter space and
the eigenvalues are then connected continuously, which has been shown
in numerical studies and
experiments~\cite{Hei99,Magunov2001a,cartarius_prl,Hernandez2006,Cartarius08a,Gutoehrlein13a,Dembowski2001,Dietz2011,Bittner2014a,Gao2015a}.

It has been shown that EP2s can be exploited for the controlled occupation
of a single quantum state \cite{Lefebvre2009a,Gilary2013a,Greaefe2013a}.
However, in these considerations the two resonances coalescing at the
exceptional point have always been assumed to be isolated from all other
states. This is very often not the case in physical systems.
\citet{Leclerc2013a} have shown that the existence of further resonances
in the vicinity of an EP can significantly influence the non-adiabatic temporal evolution and the
exchange behavior of states at an exceptional point.

In this paper we will address this question in detail. We do this by
investigating the resonances of the hydrogen atom in crossed electric and 
magnetic fields. It is especially suited for this investigation since
numerically exact calculations of resonance states are feasible, a
large number of exceptional points is known, and their properties are
clearly observable \cite{cartarius_prl,cartarius_pra,Car11b}. In
particular, examples of exceptional points with additional resonances
in their neighborhood are available. Some possess neighboring resonances
in their close vicinity, others are very isolated. In addition, a case
of two adjacent second-order exceptional points has been discovered,
of which the permutation behavior of the resonances is exactly that of
a third-order exceptional point if both exceptional points are
encircled together. This gives us the opportunity to study the
non-adiabatic state transfer at third-order exceptional points.

The remaining sections of this article are organized as follows. In
Sec.~\ref{sec:hydrogen} we introduce our system, viz.\ the hydrogen
atom in crossed static electric and magnetic fields, the numerically
accurate method to calculate the resonances of the system, and the
evaluation of the temporal evolution of occupation probabilities of
the resonances. The non-adiabatic evolution of the resonance states is
then investigated in Sec.~\ref{sec:exchange} for the case of a
second-order and a third-order exceptional point. A discussion and conclusions
are given in Sec.~\ref{sec:discussion}.

\section{Hydrogen atom in crossed electric and magnetic
  fields}
\label{sec:hydrogen}

\subsection{Resonances of the hydrogen atom}

In atomic Hartree units without relativistic corrections or finite nuclear
mass effects the Hamiltonian of the hydrogen atom in crossed static electric
and magnetic fields reads
\begin{equation}
  H = \frac12 \bm p^2 - \frac1r + \frac12 \gamma L_z
  + \frac18 \gamma^2(x^2+y^2) + fx \; ,
\end{equation}
where $L_z$ is the $z$ component of the angular momentum, and $\gamma = 
B/B_0$
and $f=F/F_0$, with $B_0 = 2.35 \times 10^5$~T 
and $F_0 = 5.14  \times 10^9$~V/cm,
are the dimensionless field strength parameters of the magnetic and electric fields, which are 
oriented along the $z$ and $x$ axis, respectively. The total energy and the
parity with respect to the $(x,y)$ plane are the constants of motion. The parity
is the only remaining symmetry of the system and is exploited in the
calculations. All subsequent studies are done in the symmetry subspace of
states with even $z$ parity. To calculate the resonances of the Hamiltonian
the complex rotation method is applied \cite{Reinhardt1982,Moiseyev1998,%
Moiseyev2011a,Delande1991,Ho1983}. By replacing the spacial coordinates
$\bm{r}$ in the Hamiltonian and the wave functions with $b^2 \bm{r}$, where
$b$ is a complex scaling parameter, we obtain a complex symmetric Hamiltonian,
in which resonances appear as discrete complex energy eigenvalues. The real
part of these complex eigenvalues represents the resonance energy, its
imaginary part the width.

With the introduction of semiparabolic coordinates a complex symmetric matrix
representation of the Schr\"odinger equation can be set up in an oscillator
basis \cite{main1994}. This leads to the generalized eigenvalue problem
\begin{equation}
  \label{eq:schroedinger-matrix}
  \bm{A}(\gamma,f) \phi = 2 |b|^4 E \bm{C} \phi \; ,
\end{equation}
where $\bm A(\gamma,f)$ is a complex symmetric matrix, $\bm C$ is a
real symmetric positive definite metric, and $E$ is the complex energy
eigenvalue. The appropriate normalization of the eigenvectors in the complex
extended system has to be done with the c-product \cite{Moiseyev2011a} and
reads for the generalized eigenvalue problem \eqref{eq:schroedinger-matrix},
$\phi_i\bm C\phi_j=\delta_{ij}$.

\subsection{Temporal evolution of the occupation probabilities
  of the resonances}

In our scheme the field strengths $\gamma(t)$ and $f(t)$ are varied
time-dependently such that closed loops are traversed in the parameter space.
This results in a time-dependent matrix $\bm{A}(t)$, and thus also
time-dependent resonance energies $E_i(t)$ and eigenstates $\phi_i(t)$. To study
the population transfer during the traversal of a closed loop around an
exceptional point we split the state by means of the spectral decomposition
into these time-dependent eigenstates $\phi_i(t)$ of the Hamiltonian, i.e.\
the expansion coefficients $a_i(t)$ define the occupation of a resonance
state $\phi_i(t)$ which is an eigenstate of the Hamiltonian with the current
field strengths $\gamma(t)$ and $f(t)$,
\begin{equation}
  \psi(t) = \sum_i a_i(t) \phi_i(t) \; .
\end{equation}
This corresponds to the instantaneous basis used in \cite{uzdin} to study
the non-adiabatic transfer in a matrix model. In the instantaneous basis the
temporal evolution of the expansion coefficients following from the
Schr\"odinger equation \eqref{eq:schroedinger-matrix} reads
\begin{equation}
  \dot{a}_i(t) = -\ii E(t) a_i(t) - \sum_j a_j(t) \phi_i(t) \bm{C}
  \dot{\phi}_j(t) \; .
  \label{eq:full_evolution}
\end{equation}
The dominant effect is the decay of the resonances, which leads to a fast
decrease of the occupation coefficients $a_i$.

For a better and more intuitive interpretation of the occupation transfer
during a path around an exceptional point we introduce weighted coefficients,
for which the overall decay of the probability amplitude is removed. They are
meant to illustrate the relative gain and loss. The weighted coefficients are
denoted by a bar and are given by
\begin{equation}
  \label{eq:weighted}
  \bar a_i = |a_i|^2 \biggl(\sum_{j=1}^N |a_j|^2 \biggr)^{\!-1} \; ,
\end{equation}
where $N$ is the total number of states taken into account.

If there were no couplings between the eigenstates, i.e.\ $\phi_i \bm
C \dot{\phi}_j=0$, all populations would evolve independently and only
the decay of the resonances with a time-dependent decay rate
$\operatorname{Im}(E_i(t))$ would be observed. This adiabatic
expectation can be formulated as
\begin{equation}
  \label{eq:adiabatic}
  \dot{a}_{\text{ad}}(t) = - \ii E(t) a_{\text{ad}}(t) \; .
\end{equation}
We use it to compare the full temporal evolution given by
Eq.~\eqref{eq:full_evolution} with the adiabatic approximation.

\section{State exchange for circles around exceptional
  points}
\label{sec:exchange}

\subsection{Second-order exceptional point}
\label{sec:ex_ep2}

\begin{figure}[tb]
  \centering
  \includegraphics[width=\linewidth]{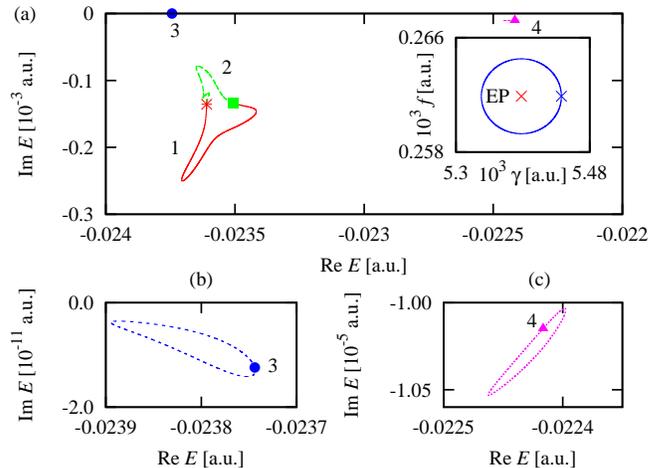}
  \caption{(Color online) (a) Two resonances interchanging their positions in
    the complex energy plane for a parameter space circle (inset) around
    the exceptional point listed in the first line of Table~\ref{tab:eps}.
    The center of the parameter space circle is identical with the exceptional
    point and $\delta = 10^{-2}$ was chosen. The initial point in the parameter
    plane and the corresponding energy values of the resonances are marked by
    symbols on the lines. The plots (b) and (c) show the dynamics of two
    nearby resonances with small imaginary parts, which are only visible as
    dots in (a).}
  \label{fig:ep2-map}
\end{figure}

In the following we study the population transfer at previously determined
exceptional points \cite{cartarius_pra}. First we observe the behavior at a
second-order exceptional point, of which the physical parameters are given in
the first row of Table~\ref{tab:eps}.
\begin{table}[tb]
  \centering
  \caption{Coordinates of some exceptional points in the spectrum of
    the hydrogen atom in crossed magnetic ($\gamma$) and electric ($f$)
    fields. All values are given in atomic Hartree units.}
  \label{tab:eps}
  \begin{ruledtabular}
  \begin{tabular}{llll}
    \multicolumn{1}{c}{$\gamma$} &
    \multicolumn{1}{c}{$f$} &
    \multicolumn{1}{c}{$\operatorname{Re}(E)$} &
    \multicolumn{1}{c}{$\operatorname{Im}(E)$} \\
    $0.005388$ & $0.0002619$ & $-0.02360$ & $-0.00015$  \\
    $0.00611$  & $0.000256$  & $-0.01593$ & $-0.00024$  \\
    $0.00615$  & $0.000265$  & $-0.0158$  & $-0.000374$ \\
  \end{tabular}
  \end{ruledtabular}
\end{table}
To encircle the exceptional point the parameters $\gamma$ and $f$ have
to be varied in a specific way. They need to perform a closed loop,
hence a good choice is a circle described by
\begin{equation}
  \label{eq:loop}
  \gamma(\varphi) = \gamma_0 (1+\delta\cos\varphi)
  \;,\quad
  f(\varphi) = f_0 (1+\delta\sin\varphi) \;,
\end{equation}
where the pair $(\gamma_0,f_0)$ represents the circle's center and $\delta$ is a
radius chosen relative to the field strengths. The trajectories of the
resonances in energy space for a relative radius $\delta=10^{-2}$ are depicted
in Fig.~\ref{fig:ep2-map}(a).
The center was chosen exactly at the exceptional point. The plot shows two
resonances interchanging their position during a traversal of the loop in
the parameter space, which is shown in the inset. Two other resonances with
a smaller modulus of the imaginary part are plotted alongside. There are
even more resonances in the vicinity with greater moduli of the imaginary
parts. Calculations taking these also into account were carried out, and it
turned out that they do not influence the result. 

\begin{figure}[tb]
  \centering
  \includegraphics[width=\linewidth]{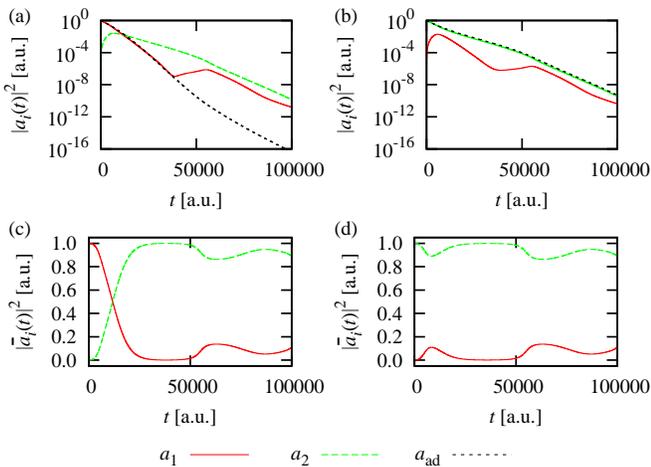}
  \caption{(Color online) Temporal evolution of the populations. Only
    the two resonances connected with the exceptional point were taken
    into account.  In (a) and (b) the actual populations are plotted
    for an initial population $a_1(0) = 1$ and $a_2(0) = 1$,
    respectively. The overall evolution is dominated by the decay
    emergent from the non-zero imaginary parts of the eigenvalues. The
    weighted coefficients according to Eq.~\eqref{eq:weighted} are
    shown for the initial condition of (a) in (c) and for that of (b)
    in (d). The results of the adiabatic approximation ($a_{\mathrm ad}$) 
    are shown for comparison.} 
  \label{fig:ep2-2res}
\end{figure}

We proceed to investigate how well the statement given in \citet{uzdin},
viz.\ that the final distribution of the populations is independent of the
initial state, is visible in the case of the hydrogen atom with its large number
of highly coupled states. Therefore we prepare the system with the full
population starting in one of the two resonances belonging to the exceptional
point. In a first step we neglect all further states, and thus effectively
reduce the Hilbert space to two dimensions. In Fig.~\ref{fig:ep2-2res}
the evolution is displayed.  The line styles of the respective
coefficients $a_i$ (excluding $a_{\mathrm{ad}}$, of course) correspond
to those shown in the map in Fig.~\ref{fig:ep2-map}.
We find that the adiabatic hypothesis is perfectly met for the case of the
population being fully prepared in $a_2$, which can be graphically
verified in Fig.~\ref{fig:ep2-2res}(b). The final state is in both cases
a majority population of the state labeled $a_2$ and a minority in that
identified with $a_1$.

\begin{figure}[tb]
  \centering
  \includegraphics[width=\linewidth]{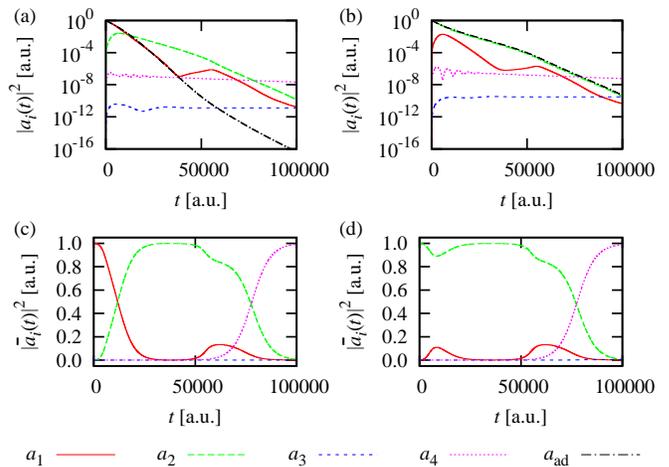}
  \caption{(Color online) Temporal evolution of the populations with
    all four resonances taken into account. In (a) the initial population
    was exclusively in the resonance described by the coefficient $a_1$ and
    in (b) only $a_2$ was initially populated. In (a) and (b) it is already
    visible that the resonances with smaller moduli of the imaginary parts
    decay much more slowly. Eventually the resonance with a smaller modulus
    of the imaginary part dominates. This is especially well visible in
    (c) and (d), in which the weighted coefficients for the cases of (a) and
    (b), respectively, are shown. The whole surviving population transfers
    into the resonance represented by $a_4$.}
  \label{fig:ep2-4res}
\end{figure}

However, this behavior changes completely if we take into account 
the resonances with smaller moduli of the imaginary parts in the 
vicinity of the two states. An example is depicted in Fig.~\ref{fig:ep2-4res},
where all four resonances shown in Fig.~\ref{fig:ep2-map} have been used
for the calculation of the temporal evolution. Due to non-adiabatic couplings
between all four resonances the states represented by $a_3$ and $a_4$ gain
in population even though no population was initially prepared in them. It is
even more surprising that resonance $a_4$ dominates in the end. In principle
one would expect the majority to end up in the resonance $a_3$ as this is a
nearly bound state with the lowest imaginary part. This is indeed what is
going to happen, but not on the time scales we used in the calculations. Since
the time is not sufficient for the occupation in $a_4$ to vanish, the stronger
coupling of that state to $a_1$ and $a_2$ decides on the final population. The
important statement holds in spite of this, viz.\ that the populations of the
states performing the exchange vanish while the populations of the states
with small imaginary parts persist.

\subsection{Importance of exceptional points for the exchange
  behavior}

\begin{figure}[tb]
  \centering
  \includegraphics[width=\linewidth]{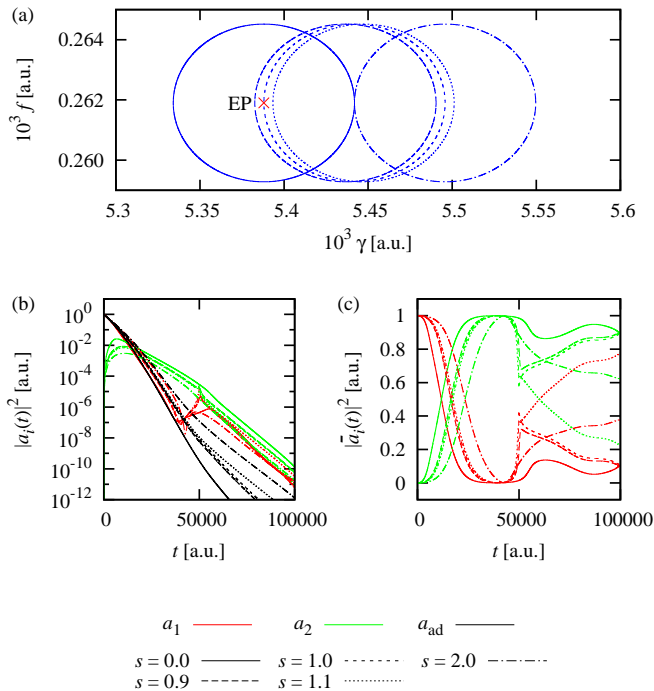}  
  \caption{(Color online) The center of the circle in parameter space
    is shifted in positive $\gamma$ direction to study the transition
    from an interchange scenario to a non-interchange case. (a) Circles for
    different values of $s$ as introduced in Eq.~\eqref{eq:loop_modified}. 
    The time evolution of the absolute (b) and weighted (c) coefficients
    demonstrate a significant qualitative change as soon as the exceptional
    point is no longer located within the parameter space loop. The line
    styles of the circles correspond to those used in the temporal
    evolution. For the non-weighted coefficients the adiabatic expectation is
    plotted alongside.}
  \label{fig:shift}
\end{figure}

We have seen that the decay rates, i.e.\ the imaginary parts of the complex
energy eigenvalues, basically determine which nearby resonance of an
initially occupied state survives at the end of the parameter space loop. This
raises the question whether the exceptional point really is important for
the exchange behavior since non-adiabatic couplings and strongly unequal decay
rates can appear without the existence of exceptional points for any pair of
resonances with similar energies as well. To address this question we move the
center of the parameter space circle as shown in Fig.~\ref{fig:shift}(a).
The first circle is, as before, exactly centered at the exceptional point. Then
it is shifted in small steps to larger values of $\gamma$ via the shift
parameter $s$. The modified circle reads
\begin{equation}
  \label{eq:loop_modified}
  \gamma(\varphi) = \gamma_0 [1+\delta ( s + \cos\varphi) ]
  \;,\quad
  f(\varphi) = f_0 (1+\delta\sin\varphi) \;.
\end{equation}
To get an intuitive insight we reduce the calculation again to the two
resonances involved in the exceptional point. This removes the overall effect
of the unavoidable transition to the slowest decaying nearby resonance and
helps us to focus on the effect of the exceptional point. 

The temporal evolution of the resonances for this case can be seen in
Fig.~\ref{fig:shift}(b) and the representation with weighted
coefficients is given in Fig.~\ref{fig:shift}(c). In all cases the
initial population was exclusively in the state labeled with the
coefficient $a_1$. This is the non-adiabatic case from above. As long
as the exceptional point is located inside the parameter space loop
there are only slight changes of the temporal evolution. In
particular, the final majority population of the coefficient $a_2$
remains unchanged. As soon as the exceptional point lies outside the
parameter space circle the evolution of the occupations changes
suddenly. One can observe that for $s=1.1$ the dominating population
is in the coefficient $a_1$, a result which agrees with the adiabatic
expectation. Since there is no longer a permutation of the resonances
even in the purely adiabatic case if the exceptional point is not
encircled, the switch of the majority population from one state to the
other is not very surprising. However, the influence of the
exceptional point is even stronger. It is additionally expressed in
the total difference of the occupation of both states. If an
exceptional point is encircled, it is much more pronounced as in the
case in which there is no exceptional point within the circle.

For an even stronger shift of the circle the difference in the final
population of $a_1$ and $a_2$ is reduced further and the majority population
switches again. This happens in a smooth way and can be traced back to the
different decay rates.

\begin{figure}[tb]
  \centering
  \includegraphics[width=\linewidth]{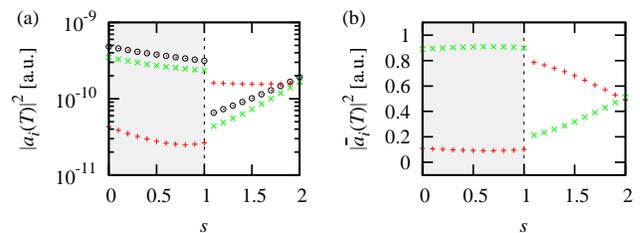}
  \caption{(Color online) Final values of the coefficients after a
    full traversal of the circle in parameter space as a function of
    the shift parameter $s$, where the absolute coefficients (a) and
    their weighted counterparts (b) are shown. Red plus symbols
    represent $a_1$, green crosses stand for $a_2$, and the black
    circles mark the adiabatic approximation. The shaded area denotes
    that the circle for this value of $s$ encircles the exceptional
    point.}
  \label{fig:final-population}
\end{figure}

The whole scenario can be understood even better with
Fig.~\ref{fig:final-population},
in which the final population of both resonances is shown in dependence on
the shift parameter $s$. The dramatic influence of the exceptional point 
becomes immediately clear due to the sudden jump of both populations at
the value $s=1$, for which the parameter space circle crosses the exceptional
point. Thus, the total behavior uncovered in Figs.~\ref{fig:ep2-2res} and
\ref{fig:ep2-4res} cannot be explained by the non-adiabatic couplings and
the different decay rates alone. The presence of the exceptional point is
essential.

\begin{figure}[tb]
  \centering
  \includegraphics[width= \linewidth]{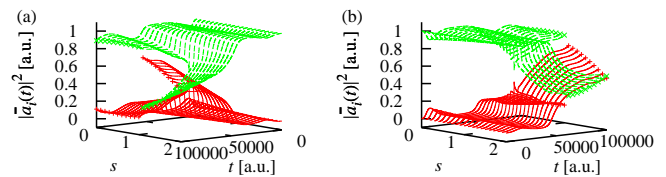}
  \caption{(Color online) Weighted coefficients $|a_i|^2$ from Fig.\
    \ref{fig:final-population} as a function of the shift parameter
    $s$ and the evolved time $t$. Two different views on the ($s$,$t$)
    plane are shown in (a) and (b). One clearly recognizes the
    dramatic change as soon as the exceptional point is within the
    parameter space circle indicating the switch from one Riemann
    surface to the other.}
  \label{fig:finalpop-3d}
\end{figure}

The relevance of the exceptional point also becomes clear in Fig.\
\ref{fig:finalpop-3d},
in which the two weighted coefficients are plotted in dependency of
the shift parameter $s$ and the evolved time $t$. It can be seen that
those paths which encircle the exceptional point lead to a dramatic
change in the occupation probabilities. For values $s \gtrapprox 1$,
i.e.\ close to the exceptional point, the switch from one Riemann
surface to the other becomes observable.  However, for larger shifts
$s\approx 2$ the non-adiabatic processes lead to a change of the final
result.

\subsection{Third-order exceptional point}
\label{sec:ex_ep3}

\begin{figure}[tb]
  \centering
  \includegraphics[width=\linewidth]{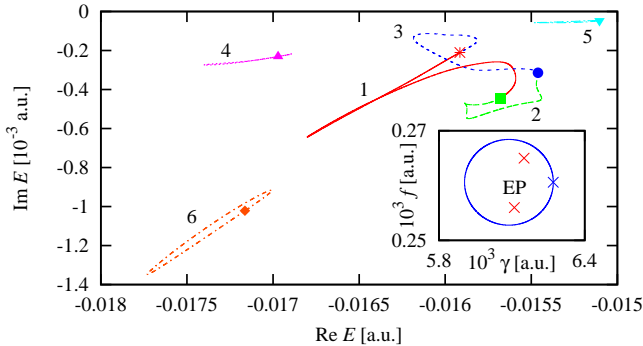}
  \caption{(Color online) Map in the energy space of a structure identical to
    that of a third-order exceptional point for a parameter space circle around
    the two exceptional points from the last two rows of Table~\ref{tab:eps}
    (see inset). The plot shows three resonances interchanging their position
    and some nearby resonances in the energy space. The energy eigenvalues
    for the initial point on the parameter space circle are marked by symbols
    on the line.}
  \label{fig:ep3-map}
\end{figure}

The hydrogen atom does not only possess second-order exceptional points. A
structure identical to that of a third-order exceptional point has also
been detected \cite{cartarius_prl}. It is uncovered by encircling the two
exceptional points in the last two rows of Table~\ref{tab:eps} at once. An
energy map of the resulting scheme of interchange is depicted in Fig.~\ref{fig:ep3-map}
for $\gamma_0 = 0.00609$, $f_0 = 0.000261$, and a radius of
$\delta = 3.0\times10^{-2}$. One can see that
all three resonances are permuted. In this case a closed loop of a single
resonance in the complex energy plane is only achieved by three circles in
the parameter space.

\begin{figure*}[tbp]
  \centering
  \includegraphics[width=\linewidth]{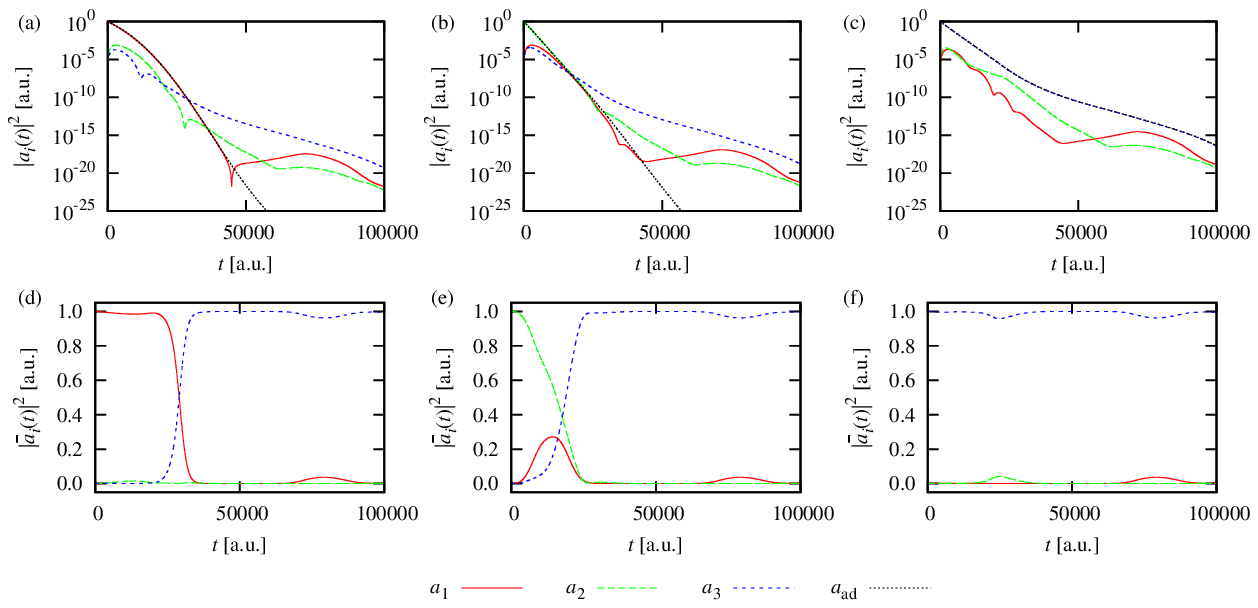}
  \caption{(Color online) Temporal evolution of the populations for a
    circle around a third-order exceptional point. The initial populations
    were set up in $a_1$ [(a) and (d)], $a_2$ [(b) and (e)], and $a_3$
    [(c) and (f)]. In the case of the initial population being prepared in
    $a_3$ the system evolves adiabatically. In all cases the non-dissipated
    population ends up in $a_3$. This can be clearly seen in the weighted
    coefficients $\bar{a}_i$.}
  \label{fig:ep3-3res}
\end{figure*}

As in the case of the second-order exceptional point we first perform
calculations which only take the three resonances connected to the EP3
structure into account. This allows us to study the non-adiabatic
temporal evolution for an unperturbed EP3. For each calculation we
prepare the initial population fully in one of these states. The
results are plotted in Fig.~\ref{fig:ep3-3res}.
To facilitate the comparison the line styles correspond to those in the map
of Fig.~\ref{fig:ep3-map}. The plots~\ref{fig:ep3-3res}(a) and (d)
correspond to the system being initially prepared in the state
$a_1$. Until about half way through the circle in parameter space the
system evolves adiabatically as can be seen in
Fig.~\ref{fig:ep3-3res}(a), i.e.\ the line of $a_1$ is exactly on top
of that for the adiabatic case. However, then the lines separate and
the state associated with $a_3$ exceeds the $a_1$ curve in
amplitude. Fig.~\ref{fig:ep3-3res}(d) shows essentially the same, but
for the coefficients weighted according to
Eq.~\eqref{eq:weighted}. From this we find that the population of the
state associated with $a_1$ is transferred into the state associated
with $a_3$, whereas the state with $a_2$ is not involved at all in the
population transfer.

Figures~\ref{fig:ep3-3res}(b) and (e) depict the situation for the initial
population being prepared in $a_2$. Even though there is some transition to
$a_1$ at first, the coefficient $a_3$ soon dominates. After $a_3$
ascends to the population leader, $a_2$ does no longer contribute to
the population. If the system is prepared with the initial population
in $a_3$ it evolves adiabatically until the end of the present cycle time
[cf.\ Figs.~\ref{fig:ep3-3res}(c) and (f)]. As was observed above for the
other cases the system also starts off adiabatically, but eventually deviates
from this behavior. Hence for the initial population in $a_1$ or $a_2$ a state
flip occurs. This is particularly visible in the weighted coefficients in
Figs.~\ref{fig:ep3-3res}(c)-(f). With these results we can extend the
statement given in \citet{uzdin} to the scenario of three permuting resonances.
Also in this case the final distribution of the populations is independent of
the initial state. In the example considered it always ends up in $a_3$.

\begin{figure*}[tbp]
  \centering
  \includegraphics[width=\linewidth]{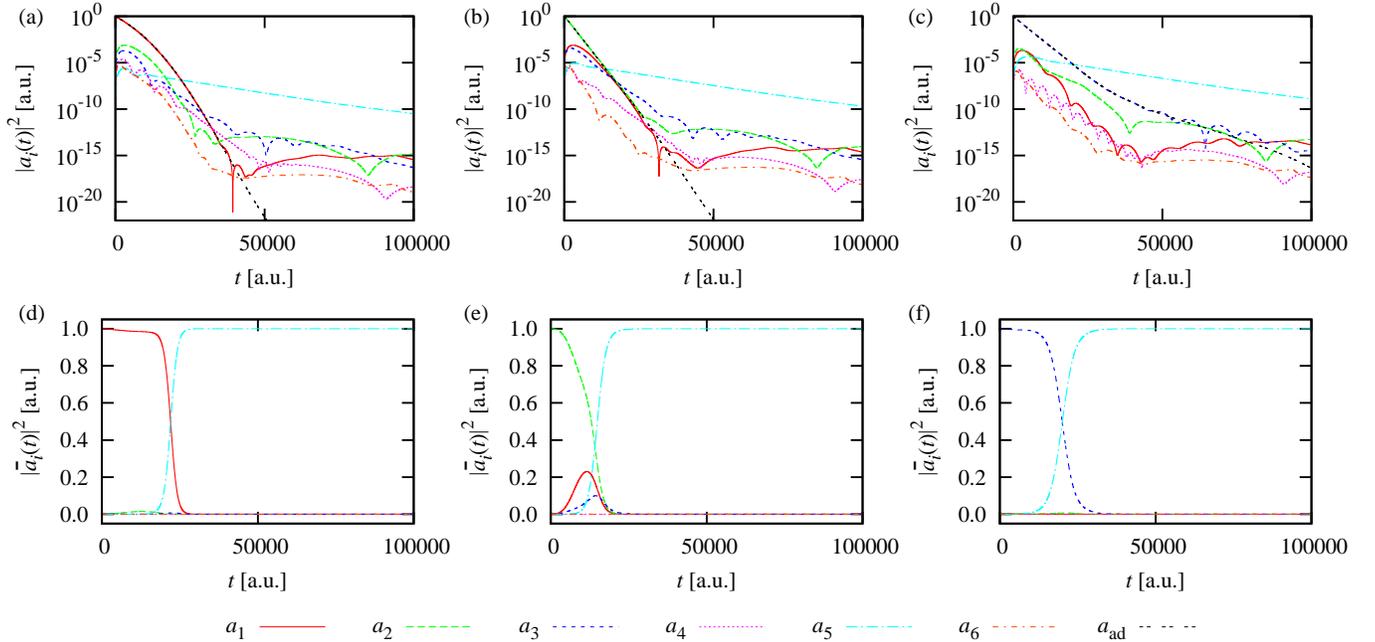}
  \caption{(Color online) Temporal evolution of the populations for the
    same parameter space circle as in Fig.~\ref{fig:ep3-3res} but with
    six resonances taken into account in the calculation. Again, the initial
    populations were set up in $a_1$ [(a) and (d)], $a_2$ [(b) and (e)],
    and $a_3$ [(c) and (f)]. The population always ends up in the resonance
    with the least modulus of the imaginary part, here $a_5$.}
  \label{fig:ep3-6res}
\end{figure*}

Of course, the three resonances forming the EP3 are not isolated in the
spectra of the hydrogen atom as we observed already for the EP2. Here the
closest three resonances have to be taken into account to obtain a realistic
temporal evolution. All of them are included in Fig.~\ref{fig:ep3-map}.
Considering the imaginary parts of the resonances during the whole parameter
space loop we have one which is strictly larger, one which is strictly
smaller, and one that lies somewhere in between that of the interchanging
resonances. The time evolution is depicted in Fig.~\ref{fig:ep3-6res}
in the same way as in Fig.~\ref{fig:ep3-3res}.

Starting from the left Fig.~\ref{fig:ep3-6res}(a) corresponds to the
initial population being prepared fully in the state represented by
$a_1$. Here, the system evolves adiabatically at first, but eventually
$a_5$ exceeds all others in terms of magnitude. This is even more
visible in the plot for the weighted coefficients shown in
Fig.~\ref{fig:ep3-6res}(d), where $a_5$ rapidly approaches unity after
about a quarter of the traversal time and then does not change for the
rest of the evolution. A similar behavior is observed for an initial
population of only the state represented by $a_2$, though $a_5$
approaches unity even faster in the weighted representation shown in
Fig.~\ref{fig:ep3-6res}(e). While the population prepared fully in the
state corresponding to $a_3$ evolved adiabatically when nearby
resonances were neglected, we again observe a transition into $a_5$ in
the extended case as can be seen in Figs.~\ref{fig:ep3-6res}(c) and
(f).

We conjecture that the transition from all other resonances to $a_5$
is induced by non-adiabatic couplings. This is certainly an effect of the
close distance between the resonance of $a_5$ and the interchanging
resonances in energy space, cf.\ Fig.~\ref{fig:ep3-map}. One could
now claim that the other resonances taken into account are also in a
close distance and should acquire a considerable occupation during the
traversal of the loop. However, this is not the case. They are even invisible
in the diagram for the weighted coefficients. This happens due to the fact
that their imaginary parts are substantially larger than that of the
resonance belonging to $a_5$, which results in a faster decay of their
population.

\section{Discussion and conclusions}
\label{sec:discussion}

In summary we were able to show that the non-adiabatic state flip at
an EP2 is also observable in the temporal evolution of occupied
resonances of the hydrogen atom in crossed electric and magnetic fields.
If only the two resonances connected to an exceptional point are taken into
account the system always ends up in the same state independent of the initial
condition as in all previous studies \cite{uzdin,Berry2011,Gilary2013a,%
  Greaefe2013a,Viennot2014a,Milburn2015a,Gilary2013a,Greaefe2013a,Leclerc2013a}
if a parameter space loop around the exceptional point is performed. However,
the spectra of the hydrogen atom always exhibit further resonances in the
vicinity of those forming the exceptional point, which can drastically
influence the final occupation \cite{Leclerc2013a}. A coupling to these states
cannot be neglected and eventually the state with the smallest decay rate
dominates. This could be verified in numerically exact calculations for the
hydrogen atom.

Even though the non-adiabatic couplings in combination with different decay
rates basically decide which resonance is occupied after a traversal of a
parameter space loop they are not the only relevant information. The temporal
evolution is strongly influenced by the presence of an exceptional point. If
it is located within the parameter space loop the difference in the final
occupation is exchanged and increases drastically.

Similar relations hold for third-order exceptional points. If couplings
to further resonances can be neglected and the traversal time is long enough
the final population always ends up in the same state and does not depend
on the initial population of the three states forming the EP3. As in the EP2
case for a realistic scenario in an atomic system further resonances have
to be considered and lead to a change of the behavior in favor of an exclusive
population of a nearby resonance in energy space with the lowest decay rate.

The calculations reported in this work clearly show that an observation of the
characteristic non-adiabatic population transfer at exceptional points will
only be possible if sufficiently isolated resonances are accessible. In an
atomic or molecular quantum system this will be a big challenge. For the
hydrogen atom we have to remark that the parameter range we used in the
calculations was chosen due to the numerical capabilities and includes
a basis with approximately 10000 states. This makes accessible
relatively low lying energies which have to be influenced by strong fields
(magnetic field $\approx 1000\,\mathrm{T}$, electric field $\approx 10^6
\,\mathrm{V/cm}$) resulting in short decay times. Even though
extremely low surviving probabilities were accepted in the calculations the
traversal time of the parameter space loop is on the order of $10^{-11}\,
\mathrm{s}$. In an experiment this could be overcome by aiming at resonances
states at higher energies, which additionally increases the probability of the
appearance of exceptional points since the density of states is higher, or by
investigating an almost equivalent system. Hydrogen-like excitons in
semiconductor structures would lower the physical parameters to accessible
values and could be studied in experiments \cite{Kazimierczuk2014a}.

%\bibliography{references}
%

\end{document}